# Quantum metric induced nonlinear transport in the hidden loop-current phase of kagome metal $RbV_3Sb_5$

Jia-Chen Shi[1], Shupeng Xu[1], Utkarsh Khandelwal[1], Nashra Pistawala[2], Luminita Harnagea[2,3], Steven J. May[4], Ritesh Agarwal[1,*]

[1]*Department of Materials Science and Engineering, University of Pennsylvania, Philadelphia, Pennsylvania 19104, United States*

[2]*Department of Physics, Indian Institute of Science Education and Research, Pune, Maharashtra 411008, India*

[3]*I-HUB Quantum Technology Foundation, Indian Institute of Science Education and Research, Pune 411008, India*

[4]*Department of Materials Science and Engineering, Drexel University, Philadelphia, Pennsylvania 19104, United States*

*riteshag@seas.upenn.edu

## Abstract

A hidden low-temperature phase with possible loop-current order has been proposed in the kagome metals $AV_3Sb_5$ ($A = \mathrm{K}$, Rb or Cs), but its experimental signatures remain subtle and indirect. Here, we use third-order nonlinear transport to probe this hidden phase in $RbV_3Sb_5$. At ~35 K, the longitudinal cubic response develops a strong kink and pronounced directional anisotropy with a strong departure from common relaxation time scaling, while the transverse cubic response acquires a magnetic field-odd component at the same temperature. Their coincident onset identifies the third-harmonic response as a sensitive marker of the low-temperature electronic reconstruction with time-reversal broken symmetry. Crucially, we find that the quantum metric quadrupole contributes directly to the longitudinal third-order response. In a loop-current charge-density-wave model, its direction selective enhancement captures the observed angular reconstruction. Our work shows how higher-order nonlinear transport translates subtle changes in electronic symmetry and quantum metric into measurable electrical signatures in quantum materials that is important for studying complex electronic phases of matter.

## Introduction

The kagome metals $AV_3\mathrm{Sb}_5$ provide a platform in which geometric frustration, non-trivial band topology, charge order, superconductivity and electronic symmetry breaking coexist[1–5]. Their kagome-derived bands host flat-band features, Dirac cones and van Hove singularities near the Fermi level[6–8], giving rise to interaction-driven electronic instabilities. A defining feature of this family is the $2 \times 2$ charge-density-wave (CDW) order[9–11], which sets in ~80–100 K and stacks the kagome plane modulation along the crystallographic $c$ axis[12–14]. Beyond this primary transition, an additional anomaly emerges ~35 K, indicating a further reconstruction of the CDW state at lower temperatures[15–25].

The microscopic origin of the 35 K state remains unsettled. Scanning tunnelling microscopy (STM)[18,26] revealed field-tunable chiral charge order, muon spin rotation ($\mu$SR)[27–29] detected enhanced internal-field distributions, and magnetotransport[15,23] showed field-switchable chiral responses. Motivated by these observations, previous studies have proposed loop-current order as a possible microscopic origin of the 35 K state[30–33]. In the proposed state, equilibrium currents circulate around bonds or plaquettes within the CDW pattern, generating orbital magnetism and breaking time-reversal symmetry (Fig. 1a). However, this interpretation remains under debate because loop-current order has not been directly imaged, and different probes have reached conflicting conclusions regarding time-reversal symmetry breaking below 35 K[34–37].

Nematicity in the CDW state (T < ~100K) presents a similar ambiguity; elastoresistance, Raman spectroscopy and other spectroscopic measurements identify a nematic transition in the CDW state[14,38,39], whereas other measurements argue that measurable anisotropy is absent in the unperturbed material[40]. Resolving the rotational and time-reversal symmetries is therefore essential for precise microscopic characterization of the quantum phase at ~35 K and also for distinguishing loop-current order from the nematic or CDW-stacking alternatives. This distinction can also be important for understanding superconductivity in $\mathrm{AV_3Sb_5}$ since the quantum metric contributes to superfluid weight in flat-band systems[41,42]. Understanding how low-temperature order modifies Bloch-state geometry may help identify which features of the CDW-ordered electronic structure are relevant to superconductivity.

Nonlinear transport provides such a symmetry-selective probe because each response tensor is strictly constrained by the crystal and magnetic point-group symmetries[43–45]. In inversion-symmetric systems, the second-order nonlinear effect is forbidden, so the leading nonlinear response can appear at third order of the driving field. This third-order response can arise from intrinsic band-geometric quantities, most notably the quantum-metric and Berry-curvature quadrupole[43,46–52]. Because this tensor encodes mirror, rotational and magnetic-symmetry constraints, third-order nonlinear transport acts as a symmetry filter for Bloch-band quantum geometry and hidden order[44,50,53].

Here, we investigate third-order nonlinear transport in $\mathrm{RbV_3Sb_5}$ as a probe of the hidden low-temperature phase (~35 K). At zero magnetic field, the longitudinal third-harmonic responses exhibit distinct anomalies near 35 K and pronounced directional dependence, whereas the transverse responses changes primarily near 100 K. To probe the magnetic character of this phase, we isolate the transverse component that is odd under reversal of an out-of-plane field. This field-odd response grows strongly with increasing field magnitude and upon cooling. Its emergence is consistent with a time-reversal symmetry broken state associated with loop-current order. Scaling analysis and microscopic calculations further indicate that conventional Drude transport and Bloch-band quantum geometry contribute differently along distinct crystal directions, with the latter entering through the quantum-metric quadrupole(QMQ). These results establish third-order nonlinear transport as a sensitive probe of the interplay between hidden symmetry breaking and quantum geometry in kagome metals.

## Results

$RbV_3Sb_5$ crystallizes in the layered hexagonal space group $P6/mmm$, with the V atoms forming a two-dimensional kagome lattice (see Methods). We study flakes exfoliated from bulk $RbV_3Sb_5$ single crystals. The flakes show metallic longitudinal resistivity, $\rho_{xx}(T)$, with a residual resistivity ratio, $\mathrm{RRR} = R(300\,\mathrm{K})/R(10\,\mathrm{K}) = 22$, indicating high sample quality (Fig. S1b). The CDW transition is also observed in the temperature derivative of the resistance, $dR/dT$ (Fig. S1c).

To probe nonlinear transport in $RbV_3Sb_5$, we fabricated a Hall-bar device from a ~200 nm thick flake on a $SiO_2$/Si substrate. A longitudinal AC electric field, $E_x^{\omega}$, was applied at 17 Hz, and the longitudinal ($V_x^{n\omega}$) and transverse voltages ($V_y^{n\omega}$), were measured for $n = 1$, 2 and 3 using lock-in amplifiers. At the fundamental frequency ($n = 1$), both voltage components scale linearly with the driving voltage. The third-harmonic voltages ($n = 3$) show a clear cubic dependence on $V_x^{\omega}$, confirming their third-order origin (Fig. 1b). By contrast, the second-harmonic signals ($n = 2$) are ~10X smaller than the third-harmonic responses and do not show quadratic scaling, consistent with the inversion symmetry of the system. As a further control, the first- and third-harmonic responses are frequency independent over the measured range, whereas the second-harmonic signal varies with frequency (Fig. S2).

We next measured $V_x^{3\omega}$ and $V_y^{3\omega}$ over a range of electric-field amplitudes at different temperatures to track the nonlinear response across the different electronic transitions. Cubic fits yielded the longitudinal ($a_x^{3\omega}$) and transverse coefficients ($a_y^{3\omega}$), which quantify the strength of the third-order response (Fig. 1c). The log–log plots (inset) make the corresponding transitions apparent in both channels. The transverse coefficient changes abruptly near the CDW transition at approximately 100 K, whereas the longitudinal coefficient shows a distinct kink near 35 K. This contrast indicates that the longitudinal and transverse third-order channels couple differently to the underlying ordered states.

To test whether conventional skew scattering accounts for the nonlinear response, we examined the normalized longitudinal coefficients as a function of the linear conductivity. Because $\sigma_{xx} \propto \tau$, skew scattering predicts[54,55], $\eta_x^{3\omega} \equiv V_x^{3\omega}/(V_x^{\omega})^3 \propto \sigma_{xx}^{\mathrm{p}}, 2 \leq \mathrm{p} \leq 3$ (Supplementary Note 1). At high temperatures, the measured response shows an approximately quadratic dependence on $\sigma_{xx}$, consistent with this expectation. Below 35 K, however, the exponent rises to ~$12$ (Fig. 1d), well beyond the range of any known scattering processes. This sharp change in scaling with $\tau$ indicates an additional longitudinal contribution that emerges in the low-temperature phase (discussed later).

To further characterize the third-order response, we measured $V_x^{3\omega}$ along different in-plane directions using twelve electrodes arranged in a circular geometry (Fig. 2a). Four-probe configurations allowed the driving field to be rotated while the corresponding longitudinal response was recorded. We first summarized the angular dependence by extracting the cubic fitting parameter $a_x^{3\omega}$ as a function of $\theta$ at 10 and 60 K (Figs. 2b, c). At 60 K, a single maximum is observed at ~$120°$ within the measured angular range. At 10 K, an additional enhancement appears near $\theta = 30°$, while the $\theta = 90°$ response remains weak at both temperatures. No comparable temperature induced changes in angle-resolved data are observed in the first-harmonic resistance (Fig. S4), highlighting the sensitivity of the third-harmonic response to the low-temperature electronic reconstructions.

Representative traces of $V_x^{3\omega}$ versus $V_x^{\omega}$ are shown for selected angles at 10 K and 60 K. in Figs. 2c–e. At 60 K, all measured directions exhibit clear cubic scaling. At 10 K, however, the response exhibits an anisotropic pattern, different from that at 60 K. One direction is suppressed to near the noise level, whereas the other directions retain cubic-dependent responses, suggesting in-plane sixfold rotational-symmetry breaking.

To follow the evolution of the directional anisotropy, we also measured $\eta_x^{3\omega}$ along three representative directions as a function of temperature (Figs. 2f–h). Upon cooling below ~35 K, the $\theta = 60°$ response increases rapidly and follows the grey dashed reference curve. This curve was obtained by digitizing and normalizing the second-harmonic nonlinear transport data previously reported in $\mathrm{CsV_3Sb_5}$[15], providing a visual reference for the low-temperature crossover trend. The same trend has been reported by several other studies, including the anomalous Nernst effect and electrical magnetochiral anisotropy measurements[17,18,25]. By contrast, the $\theta = 90°$ response initially increases upon cooling, consistent with a conventional scattering contribution, but decreases below 35 K. These contrasting trends link the direction-dependent nonlinear response to the low-temperature phase emerging near 35 K.

Although the angular anisotropy reveals the altered rotational symmetry of the low-temperature phase, it does not by itself determine the associated magnetic symmetry. In order to probe whether the low-temperature phase hosts a field-coupled, time-reversal-breaking order parameter, we applied an out-of-plane magnetic field, $B \parallel z$, and measured the field-odd first- and third-harmonic transverse responses. At 15 K, the normalized transverse response $\eta_y^{3\omega} = V_y^{3\omega}/(V_x^{\omega})^3$ varies strongly with magnetic field (Fig. 3a). We isolated its odd-in-field component, $\eta_y^{3\omega,\mathrm{odd}}$, at several representative temperatures (Fig. 3b). The response remains weak above 40 K which then increases steadily with cooling and with increasing magnetic field.

We performed the same antisymmetrization for the first-harmonic response $\eta_y^{1\omega} = V_y^{1\omega}/V_x^{\omega}$, with the corresponding data presented in Fig. $\mathrm{S5}$. For a direct comparison, Fig. 3c compares the normalized first- and third-harmonic B-field-odd responses at $B = 5.2\ \mathrm{T}$. The third-harmonic response emerges below 40K and shows a substantially larger normalized change in comparison to the first-harmonic response, which remains nearly temperature independent above 40K but decreases slightly upon cooling. This contrast highlights the enhanced sensitivity of nonlinear transport to this electronic transition.

The emergence of a strong magnetic field-odd third-harmonic response below 40 K is clearly related to the low-temperature electronic reconstruction. A comparable temperature scale appeared in μSR measurements of charge-order $\mathrm{RbV_3Sb_5}$, which revealed a second steep increase in the muon-spin relaxation rate near 40 K[28]. Field-switchable chiral transport was also observed to emerge below ~35 K in $\mathrm{CsV_3Sb_5}$[15]. Together, these observations motivate loop-current order as one possible interpretation[32,56], in which time-reversed current patterns couple oppositely to an out-of-plane magnetic field. Although the field-odd response does not uniquely establish loop currents, it identifies third-order Hall transport as a complementary probe of the magnetic and chiral phenomenology associated with the hidden low-temperature phase.

## Symmetry analysis

The third-order response provides a symmetry-sensitive probe of the low-temperature phase. Whether $\mathrm{AV_3Sb_5}$ retains sixfold rotational symmetry below the CDW transition (<100 K) remains debated[16,24], with different experiments reporting either nematicity or sixfold symmetric responses. In our experiments, the angle-resolved third-order measurement directly probes the in-plane rotational symmetry. Although no electrode pair was aligned with a known crystallographic axis, the twelve electrode geometry fixes the relative angles between measurement directions and therefore resolves rotational anisotropy. We define $\theta_0$ as the angle between the applied electric field and a reference crystal axis (Fig. 4a). For a $C_6$-symmetric state, the time-reversal-even third-order conductivity yields an isotropic longitudinal response and a vanishing transverse response: $J_{\parallel}^{3\omega} = \sigma_{xxxx}|E(\omega)|^3, \quad J_{\perp}^{3\omega} = 0$. The longitudinal response is therefore independent of the in-plane electric-field direction. This expectation is consistent with the negligible transverse signal at temperatures above the CDW transition, where the sixfold rotational symmetry is preserved. The abrupt emergence of a transverse response below the CDW transition indicates that the ordered state relaxes this symmetry constraint and allows additional third-order tensor components (see details of the tensor components and symmetry constraints in Supplementary Note 2).

Our measurements instead show distinct temperature dependences along different directions, demonstrating that the low-temperature nonlinear response (<100 K) is incompatible with a simple $C_6$-symmetric state. Since no reproducible second-harmonic signal was detected, it is consistent with preserved in-plane $C_2$ symmetry. Since the flakes are dry-transferred onto prepatterned electrodes, substrate coupling may introduce some residual anisotropic strain. Even weak strain can act as a symmetry-breaking perturbation that selects or amplifies a $C_2$ electronic response, as demonstrated in a study of $\mathrm{CsV_3Sb_5}$[24]. It may also pin one of the three symmetry-equivalent $C_2$ orientations. In our experiments, the electrode directions are not independently registered to the crystallographic axes, hence we can resolve the relative anisotropy but cannot identify the selected crystallographic direction. Repeated temperature sweeps on fixed Hall-bar devices reproduced the kink near 35K and the subsequent evolution of the strong longitudinal third-harmonic response, demonstrating that the projected response is reproducible and consistent with a pinned $C_2$ orientation from intrinsic symmetry breaking.

We therefore analyze the third-order conductivity using tensor components allowed by $C_2$ symmetry. An electric field $E^{\omega}$ applied at angular frequency $\omega$ generates a third-harmonic current density by a fourth-rank conductivity tensor $\sigma_{abcd}$, via $J_a^{3\omega} = \sigma_{abcd}E_b^{\omega}E_c^{\omega}E_d^{\omega}$, where $a$, $b$, $c$ and $d$ denote spatial directions and repeated indices are summed. Because the three driving fields are identical, only the part of $\sigma_{abcd}$ that is symmetric under permutations of its last three indices is physically observable. We therefore adopt the symmetrized convention $\sigma_{xxyy} = \sigma_{xyxy} = \sigma_{xyyx}$ and $\sigma_{yyxx} = \sigma_{yxyx} = \sigma_{yxxy}$, so that each set of equivalent components enters through a single independent coefficient. With this convention, the longitudinal conductivity projection is,

$$\sigma_{\parallel}^{3\omega}(\theta_0) = \cos^4\theta_0\,\sigma_{xxxx} + 3\cos^2\theta_0\sin^2\theta_0\left(\sigma_{xxyy} + \sigma_{yyxx}\right) + \sin^4\theta_0\,\sigma_{yyyy}.$$

Within the Boltzmann framework, the third-order conductivity can be decomposed into Drude, QMQ contributions (details in the Supplementary Note 3). The band energy normalized quantum metric is[57–59],

$$G_n^{ab} = \sum_{m\neq n} \frac{A_{nm}^a A_{mn}^b + A_{nm}^b A_{mn}^a}{\varepsilon_n - \varepsilon_m},$$

where $A_{mn}^a$ is the interband Berry connection and $m$ and $n$ label the bands. The quantity $\partial_{k_a}\partial_{k_b} G_n^{ab}$ defines the corresponding normalized QMQ. These geometric terms can contribute to the third-order response even when inversion symmetry suppresses the second-order signal.

The different mechanisms can be distinguished in part through their relaxation-time dependences. The Drude and quantum-metric contributions scale as $\tau^3$ and $\tau$, respectively, and skew scattering extends this range to $\tau^4$ at most[58,59]. Because $\sigma_{xx} \propto \tau$, each of these mechanisms predicts a power-law relation between $\eta_x^{3\omega}$ and $\sigma_{xx}$ with an exponent of at most 3, provided that the band-structure prefactors remain temperature independent. This expectation is met above 35 K, where the measured exponent is approximately 2. The exponent of ~12 measured below 35 K therefore cannot be assigned to any of these mechanisms. Instead, it marks a breakdown of relaxation-time scaling: below the 35 K transition, the growing order parameter reconstructs the bands and their geometric quantities, so the third-order response is no longer a function of $\sigma_{xx}$ alone. This breakdown indicates that the enhancement reflects an order parameter-driven change of the intrinsic band geometry rather than a change only in scattering strength.

Taken together, the angle-dependent measurements identify a low-temperature response incompatible with $C_6$ symmetry, while the *B*-field odd transverse signal is consistent with time-reversal-symmetry breaking below 35 K. These complementary spatial and magnetic symmetry constraints, together with the anomalous temperature scaling, motivate testing whether a microscopic loop-current model can reproduce the observed longitudinal anisotropy and transverse response.

**Microscopic model**

To model the loop-current order (<35 K) within a microscopic framework, we calculate the third-order conductivity using a single-orbital tight-binding model on the kagome lattice. The unmodulated Hamiltonian is $H_{\mathrm{tb}} = -t\sum_{\langle i,j\rangle}\left(c_i^\dagger c_j + \mathrm{h.c.}\right) - \mu\sum_i n_i$, where $t$ is the nearest-neighbour hopping amplitude and $\mu$ is the chemical potential. To describe the $2a_0 \times 2a_0$ ordered state, we add a triple-$\boldsymbol{Q}$ modulation of the antisymmetric bonds[31,33],

$$H_{\mathrm{cdw}} = \sum_{(\alpha\beta\gamma),\boldsymbol{r}} \rho_\gamma \cos\left(\boldsymbol{Q}_{\mathrm{c}}^{\gamma}\cdot\boldsymbol{r}\right) X_{\alpha\beta}^{-}(\boldsymbol{r}) + \mathrm{h.c.}$$

where $\rho_\gamma$ is the complex CDW order parameter associated with $\boldsymbol{Q}_{\mathrm{c}}^{\gamma}$. Its real part describes the bond amplitude modulation, whereas its imaginary part generates circulating currents and breaks time-reversal symmetry. We evaluate $H = H_{\mathrm{tb}} + H_{\mathrm{cdw}}$ in a 12-site supercell using $\boldsymbol{\rho} = (0.01 +$ $i\rho_I, 0.02, 0.02)$. The unequal real amplitudes encode a nematic triple-$\boldsymbol{Q}$ bond order in which one modulation direction is inequivalent to the other two. This time-reversal-even imbalance reduces the rotational symmetry from $C_6$ to $C_2$ already at $\rho_I = 0$, consistent with the twofold symmetry reported within the charge-ordered state and with the anisotropic response measured at 60 K (Fig. 2b) (calculation details in Supplementary Note 4). The parameter $\rho_I = \mathrm{Im}(\rho_1)$ controls the loop-current amplitude, which breaks time-reversal symmetry and further reorganizes the anisotropic response below 35 K. To relate the parameter sweep to the experiment, we assume that $\rho_I$ increases upon cooling below 35 K, as expected for the amplitude of an emerging order parameter.

The reconstructed band structure for $\rho_I = 0.02$ is shown in the reduced Brillouin zone in Fig. 4b. The loop-current modulation opens gaps near the $M$ points. To resolve the associated quantum geometry, we calculate the normalized QMQ $\partial_x \partial_x G^{xx}$ for the band highlighted in red. Its momentum-space distribution is shown in Fig. 4c. We next evaluate all the tensor components allowed by $C_2$ symmetry and combine them using the longitudinal projection in Eq. (1). At $\theta_0 = 0°$, both the Drude and QMQ terms contribute to the longitudinal response (detailed expression in Supplementary Note 3). As $\rho_I$ increases, the normalized Drude term decreases, whereas the QMQ term is enhanced (Fig. 4d). To compare the calculations with the angle-resolved measurements, we evaluate the longitudinal projection $\sigma_x^{3\omega}(\theta_0)$. The Drude and QMQ terms exhibit distinct angular profiles (Figs. 4e). The Drude response is maximal near $0°$,whereas the QMQ response peaks approximately $90°$ away. As $\rho_I$ increases, the enhanced quantum metric contribution therefore adds a second angular maximum to the Drude-dominated profile. This superposition is consistent with the additional low-temperature maximum observed experimentally (Fig. 2b).

We quantified this evolution by decomposing the $a_x^{3\omega}$ data from Fig. 2b at 10 K and 60 K according to $a_x^{3\omega}(\theta, T) = C_T + A_D D_T(\theta_0 + \phi_0) + A_{QMQ} Q_T(\theta_0 + \phi_0)$. Here, $D_T$ and $Q_T$ are the calculated Drude and QMQ angular profiles, respectively, and $C_T$ is a temperature-dependent constant background. We used $\rho_I = 0.02$ at 10 K and $\rho_I = 0$ at 60 K. A common phase offset, $\phi_0 = 50.3°$, was applied to both temperatures and both contributions. Within this normalization, the 60 K response has a larger Drude amplitude, with $A_{D,60\mathrm{K}} = 3.58 \times 10^5$ and $A_{QMQ,60\mathrm{K}} = 1.97 \times 10^5$. At 10 K, the second maximum near 30° is accompanied by a larger QMQ amplitude, with $A_{QMQ,10\mathrm{K}} = 1.29 \times 10^6$ and $A_{D,10\mathrm{K}} = 1.07 \times 10^6$. Because the same phase offset is used at both temperatures, the evolution arises from changes in the relative amplitudes rather than an arbitrary rotation of the angular profiles. Within this decomposition, the fitted QMQ amplitude increases upon cooling, with its angular contribution reaching a maximum between the measured $30°$ and $60°$ directions. This result supports a direction-selective enhancement of the low-temperature QMQ response, although its quantitative assignment remains model dependent.

The contrasting temperature dependences in Figs. 2f–h provide further evidence on the origin of the direction-selective nonlinear response. Conventional Drude response cannot produce them on its own, and its temperature dependence is set by the relaxation time, so every direction should evolve in the same way. Calculations based on the tight-binding model instead yields a direction-selective evolution. As $\rho_I$ grows, the QMQ is enhanced when the electric field is parallel to $a_1$ and suppressed when it is perpendicular, whereas the Drude weight is reduced at all angles (Figs. 4e).

Along directions where the loop-current order suppresses both the Drude and QMQ contributions, their combined reduction can overcome the relaxation time driven increase, which should therefore first rise upon cooling and then turn down below the ~35 K transition. This is precisely the behavior observed at $\theta = 90°$, where the response increases before approaching 35 K and is suppressed below it (Fig. 2h). The low-temperature electronically ordered state thus enhances the nonlinear response along some directions while weakening it along others, and the model suggests a loop-current-enhanced QMQ as a source of this anisotropic reorganization.

## Conclusion and outlook

Third-order nonlinear transport uncovers an electronic reconstruction in $\mathrm{RbV_3Sb_5}$ near 35 K. Below this temperature the longitudinal response turns strongly anisotropic and stops obeying relaxation-time scaling, while the transverse response picks up a B-field-odd component at the same onset. A microscopic loop-current CDW model captures both observations: a nematic bond imbalance lowers the rotational symmetry from $C_6$ to $C_2$, and the imaginary part of the bond order breaks time-reversal symmetry and enhances the QMQ along the angular phase we observed experimentally. The tight-binding loop-current model illustrates how loop-current order can enhance the QMQ and account for the observed direction selective nonlinear response. However, developing a unified theory that incorporates the realistic multiorbital band structure[7], electronic correlations[17,18], and magnetic-field coupling[15,19,23] therefore remains an important open challenge.

Comparisons across the $A\mathrm{V_3Sb_5}$ family, together with local and domain-sensitive magnetic probes, should clarify whether the low-temperature reconstruction and B-field-odd response originate from loop-current order. Since the quantum metric can contribute to superfluid weight in flat-band systems, establishing how this phase alters Bloch-state geometry may also illuminate its relationship to superconductivity[41,42]. Beyond kagome metals, higher-order nonlinear transport offers a route for investigating hidden, multipolar and compensated magnetic phases whose linear-transport signatures may be weak[43,44,46,53]. Our findings position third-order nonlinear transport as a sensitive probe of how emergent order reshapes electronic symmetry and Bloch-band quantum geometry.

## Methods

**Single-crystal growth**. High-quality single crystals $\mathrm{RbV_3Sb_5}$ were grown from a flux composed of RbSb–$\mathrm{RbSb_2}$ eutectic mixture and excess Sb, following previously reported procedures[2,60–62]. The precursors were handled in an argon-filled glovebox, loaded into an alumina crucible and sealed under vacuum in a double quartz ampoule. The mixture was held at 550 °C for 48 hours and 1000 °C for 16 hours, followed by cooling to 600 °C at 2.5 °C/hour. Single crystals with lateral dimensions of a few millimeters were mechanically separated from the flux (Fig. S7a – inset).

X-ray diffraction and Laue measurements confirmed that the crystal surface is parallel to the crystallographic $ab$ plane (Supplementary Fig. S6a). SEM–EDX measurements showed that crystals from the same batch were homogeneous and stoichiometric within an experimental uncertainty of 1–2 at.% (Supplementary Fig. S6b). Magnetic susceptibility measured under zero-field-cooled and field-cooled conditions, with a 1T field applied parallel to the $ab$ plane, revealed a step-like anomaly near 102 K associated with the onset of CDW order (Supplementary Fig.S6c).

**Device fabrication.** Hall-bar and circular twelve-electrode patterns were defined by conventional photolithography, followed by the deposition of 10 nm Ti and 10 nm Au. Mechanically exfoliated $\mathrm{RbV_3Sb_5}$ flakes were then dry transferred onto the prepatterned electrodes.

**Linear and nonlinear transport measurements.** Transport measurements were performed in a Quantum Design Physical Property Measurement System using standard lock-in techniques. The ac excitation, $V_{\mathrm{x}}(\omega) = V_0 \sin(\omega t)$, was supplied by the internal oscillator of an SR860 lock-in amplifier (Stanford Research Systems), and phase-matched SR860 amplifiers simultaneously recorded the longitudinal and transverse voltages at the first, second and third harmonics, $V_{x/y}^{n\omega}$ with $n = 1$, 2 and 3.

**Acknowledgments** We thank Mingsheng Tian for the valuable discussions and helpful suggestions. R.A. was supported by the Office of Naval Research (grant No. N000142512360) and the National Science Foundation (grant No. DMR-2508192). S.J.M. (assistance with transport measurements) was supported by the U.S. Department of Energy (DOE), Office of Science, Basic Energy Sciences (BES), under Award No. DE-SC0024204. L.H. acknowledges funding from the National Mission on Interdisciplinary CyberPhysical Systems (NM-ICPS) of the Department of Science and Technology, Government of India, through the I-HUB Quantum Technology Foundation, Pune.

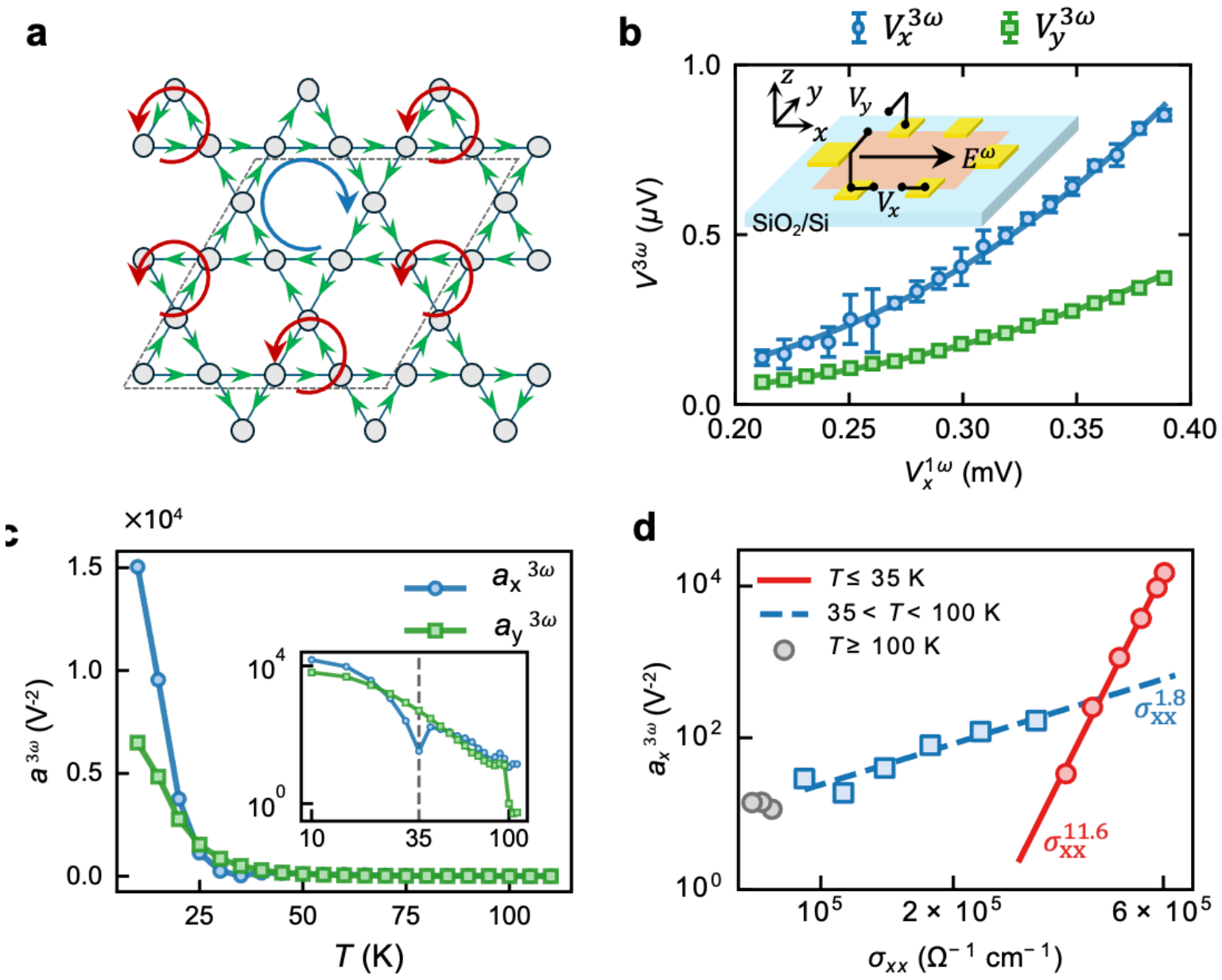

**Figure 1. Observation of third-order nonlinear transport in $RbV_3Sb_5$. a**, Schematic of the loop-current state in the 2×2 CDW phase. Arrows indicate the directions of currents, and the dashed gray line marks the 2×2 CDW unit cell[31,33]. **b,** $V_x^{3\omega}$ and $V_y^{3\omega}$ as functions of $V_x^{\omega}$ at $T = 10$ K. Solid lines show cubic fit. Insert shows schematic of the Hall-bar device geometry and electrical transport measurement configuration. **c,** Temperature dependence of the cubic fitting parameter, $a^{3\omega}$, extracted from the fits to $V_x^{3\omega}$ and $V_y^{3\omega}$. Insets show the corresponding log–log plots. **d,** Log–log plot of the longitudinal response as a function of conductivity. Gray points correspond to $T \geq 100$ K, blue squares to $35 < T < 100$ K, and red circles to $T \leq 35$ K. Dashed and solid lines represent fits to the data in the different temperature regimes.

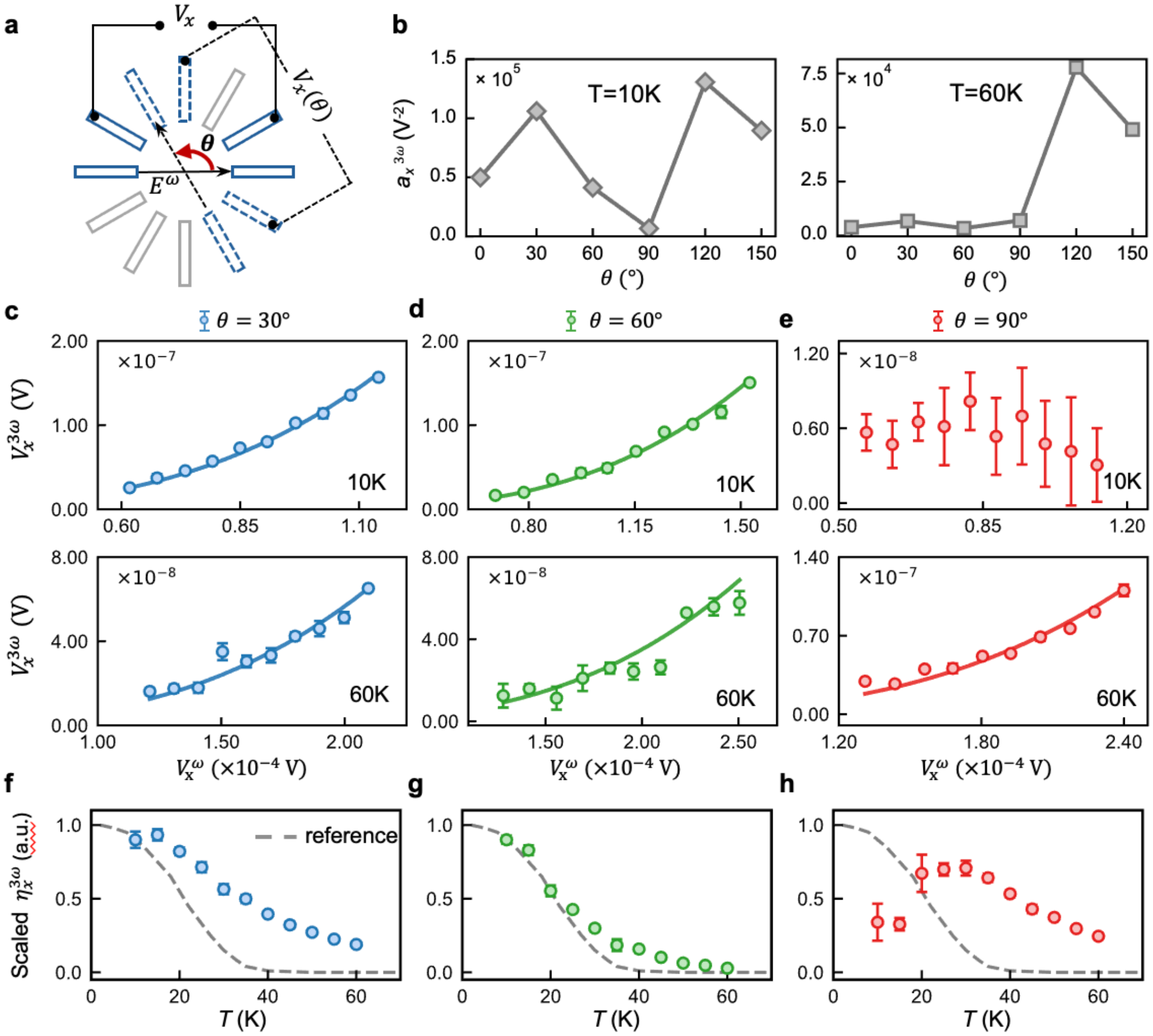

**Figure 2. Angular dependence of the third-order longitudinal response in $RbV_3Sb_5$. a,** Schematic of the circular twelve electrode device used to probe angle-dependent transport. By rotating the measurement geometry by an angle $\theta$, signals along different lattice directions are measured. **b,** Cubic fitting parameter, $a_x^{3\omega}$, as a function of $\theta$ at $T = 10$ K and $T = 60$ K. **c–e,** $V_x^{3\omega}$ as a function of $V_x^{\omega}$ at $T = 10$ K (upper panels) and $T = 60$ K (lower panels) for $\theta = 30°$, $60°$ and $90°$. Solid lines show cubic fits. **f–h,** Ratio $V_x^{3\omega}/(V_x^{\omega})^3$ as a function of temperature for $\theta = 30°$, $60°$, and $90°$. The dashed line shows data digitized from the curve in Fig. 2c of ref. [*15*] and normalized for comparison, serving as a guide to the temperature scaling below $T \approx 30$ K.

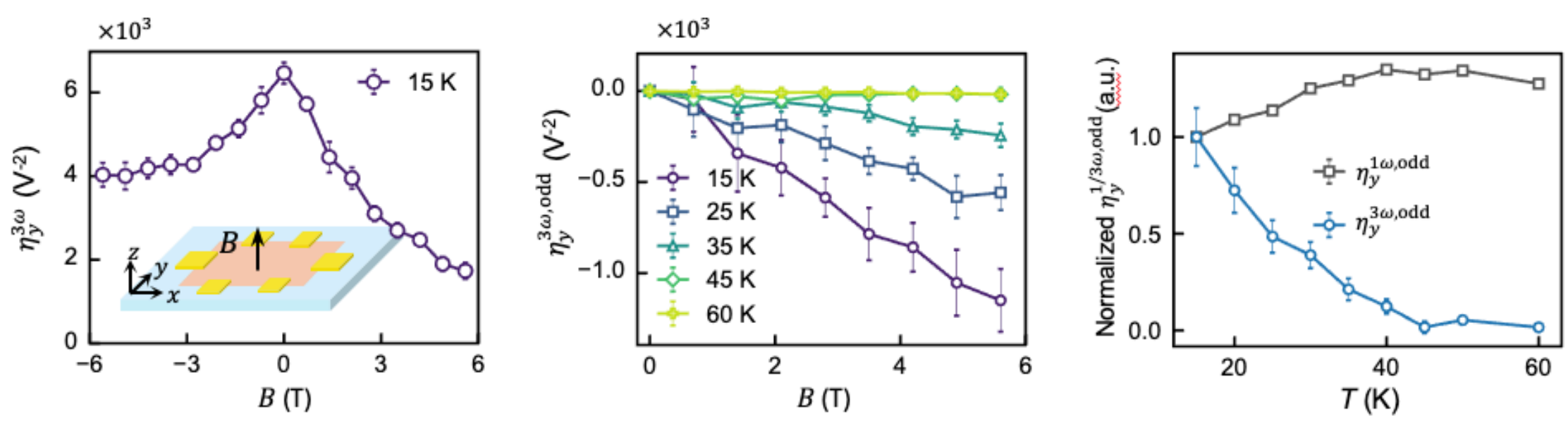

**Figure 3. B-field odd third-order Hall response below 35 K. a,** Third-order nonlinear Hall response $\eta_y^{3\omega} = V_y^{3\omega}/(V_x^{\omega})^3$ as a function of the out-of-plane magnetic field $B$ at $T = 15$ K. **b,** Odd-in-field component $\eta_y^{3\omega,odd}$ as a function of $B$ at various temperatures. **c,** Temperature dependence of the B-field odd first- and third-harmonic transverse responses, $\eta_y^{1\omega,odd}$ and $\eta_y^{3\omega,odd}$, respectively, at $B$ = 5.2 T.

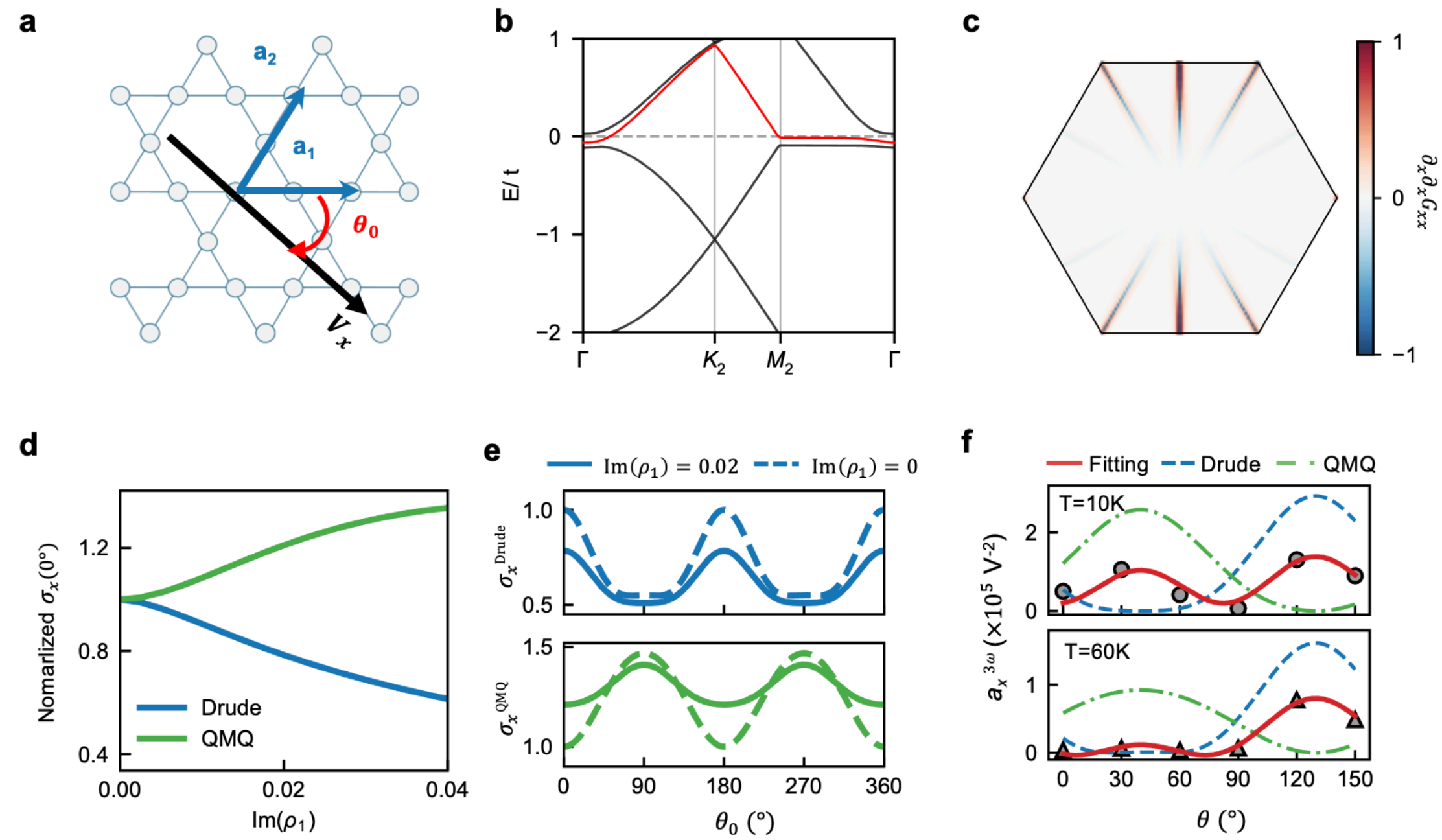

**Figure 4. Loop current enhanced quantum metric quadrupole in the kagome model. a,** Kagome lattice and primitive translation vectors $a_1$ and $a_2$. The angle $\theta_0$ is defined between the applied electric field and the $a_1$ axis. **b,** Electronic band structure for $\rho = (0.01 + 0.02i,\ 0.02,\ 0.02)$. See text for details. The band highlighted in red is used to evaluate the momentum-resolved quantity in **c**. **c,** Normalized momentum-space quantum metric quadrupole (QMQ) distribution of $\partial_x \partial_x G^{xx}(k_x, k_y)$ within the first Brillouin zone. **d,** Drude and QMQ contributions to the third-order longitudinal response at $\theta_0 = 0°$ as functions of $Im(\rho_1)$. Each contribution is normalized by its value at $Im(\rho_1) = 0$. **e,** Angular dependence of the normalized Drude (**upper panel**) and QMQ (**lower panel**) contributions for $Im(\rho_1) = 0$ and $0.02$. **f,** Decomposition of the angular nonlinear response into Drude and QMQ contributions. Grey symbols reproduce the experimental $a_x^{3\omega}$ data from Fig. 2b at 10 K (upper panel) and 60 K (lower panel). Red curves show the overall fits. Blue dashed and green dash-dotted curves denote the fitted Drude and QMQ contributions respectively.